# COMPACT CODING USING MULTI-PHOTON TOLERANT QUANTUM PROTOCOLS FOR QUANTUM COMMUNICATION

Rasha El Hajj, Pramode Verma and Kam Wai Clifford Chan

Telecommunications Engineering Program, School of Electrical and Computer Engineering, University of Oklahoma-Tulsa, USA

*ABSTRACT*

*This paper presents a new encryption scheme called Compact Coding that encodes information in time, phase, and intensity domains, simultaneously. While these approaches have previously been used one at a time, the proposed scheme brings to bear for the first time their strengths simultaneously leading to an increase in the secure information transfer rate. The proposed scheme is applicable to both optical fibers and free space optics, and can be considered as an alternative to polarization coding. This paper applies the proposed compact coding scheme to multi-photon tolerant quantum protocols in order to produce quantum-level security during information transfer. We present the structure of the proposed coding scheme in a multi-photon environment and address its operation.*

## 1. INTRODUCTION

Secrets exist in every aspect of our lives. The way civilizations have protected their secrets is as interesting as the secrets themselves. For thousands of years, kings, queens and generals have been aware of the threat of their messages falling in the wrong hands and exposing precious secrets to the enemies [1]. This desire for secrecy motivated the development of methods to code messages in such a way that only the intended recipient can read them. Until today, the centuries-old battle between code makers and code breakers has not reached an end. Every now and then, a security-related scandal about confidential information being revealed pops up. In April 2014, the Heartbleed bug, a serious vulnerability in the popular Open SSL cryptographic software library, was announced to the public [2]. Heartbleed bug allowed hackers to extract data from massive databases containing usernames, passwords and other sensitive information. Affected websites and online services had to update their servers, and users were asked to change their passwords to neutralize the vulnerability.

Contemporary cryptographic schemes are based on the use of one-way mathematical functions [3]. Their security will likely be compromised with the development of quantum computers which have the potential to carry out mathematical operations in record time. Such limitations provide the greatest incentive for quantum cryptography which is unconditionally secure.
Quantum Key Distribution (QKD) is the first and most successful quantum information processing protocol. It allows two parties Alice and Bob to securely share a string of data (key) even in the presence of an adversary, Eve. BB84, proposed by Bennett and Brassard in 1984, has been widely heralded as the first QKD protocol with commercial potential. However, it suffers from the problem of single photon generation and the attendant distance and speed limitations.

The multi-photon tolerant quantum protocols, used in this paper, mitigate the problems associated with single photon based protocols.

Most contemporary quantum cryptographic protocols are based on polarization coding due to the ease of implementation. Polarization nevertheless is subject to variations caused by a large number of ambient factors which are largely uncontrollable. This necessitates examining alternative coding schemes like phase and time coding. While these approaches have been used one at a time, this paper takes advantage of several properties of photons to encode information.
The paper is organized as follows. In section 2, multi-photon tolerant quantum protocols are described. In section 3, time coding, phase coding and intensity coding schemes are addressed. The new coding scheme is proposed in section 4. The application of the proposed coding scheme to multi-photon tolerant quantum protocols, i.e., the three stage protocol and the single stage protocol, is presented in section 5. In section 6, an analysis of the system and possible future work are tackled. Section 7 concludes the paper.

## 2. MULTI-PHOTON TOLERANT QUANTUM PROTOCOLS

Current implementations of QKD are based on the BB84 protocol. However, BB84 requires using no more than a single photon per pulse. This imposes distance and speed limitations when the channel loss and the inefficiency of single photon generation are taken into account. The limitations of single photon based approaches are overcome with multi-photon tolerant quantum cryptography protocols discussed in the following subsections.

### 2.1. Three Stage Protocol

First proposed by Subhash Kak in 2006 [4], the three stage protocol securely transmits data over a telecommunication channel. Alice sends to Bob photons in arbitrary states of polarization unlike the BB84 where the photons should be one of only four possible states. Alice and Bob use secret unitary transformations $U_A$ and $U_B$ that should be commutative, i.e., $U_A U_B = U_B U_A$ [4].

Let $X$ be the state that Alice wishes to transfer to Bob. The steps of the three stage protocol, shown in Fig. 1, can be described as follows. Alice applies $U_A$ on qubit $X$ obtaining $U_A(X)$, and sends it to Bob. Bob applies $U_B$ on the received qubit to get $U_B U_A(X)$, and sends it back to Alice. Alice negates her transformation by applying $U_A^\dagger$ (transpose complex conjugate of $U_A$) on the received qubit to obtain $U_A^\dagger U_B U_A(X) = U_B(X)$ and sends it to Bob. Bob negates his transformation by applying $U_B^\dagger$ on the received qubit to obtain $U_B^\dagger U_B(X) = X$, which is the information sent by Alice.

From the point of view of practical implementation, the advantage of the three stage protocol is that information exchange is not restricted to transmission based on a single photon. The security of the three stage protocol depends on the fact that Alice and Bob independently transform their photonic streams. These transforms can be changed at any time and do not need to be communicated to the other party. The three-stage protocol is secure against the photon number splitting (PNS) attack as long as the number of photons in the beam launched by Alice is below a certain threshold number [5, 6].

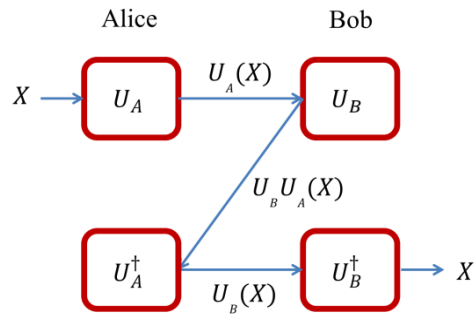

Figure 1. The three stage protocol

## 2.2. Single Stage Protocol

In the three stage protocol, information travels through the channel three times. The single stage protocol, proposed by Thomas in 2007 [7], utilizes the channel once instead of three times as shown in Fig. 2. Knowing the value of $\theta$, Bob can apply the transformation $U_A^\dagger$ to get $X$ such that $U_A^\dagger U_A(X) = X$. If Eve captures the transmitted message $U_A(X)$, unless she knows the value of $\theta$, she cannot recover the original message $X$.

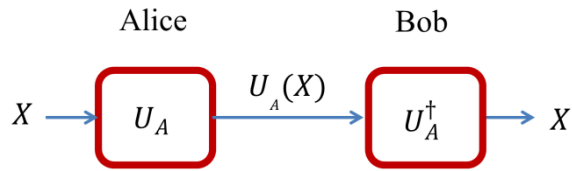

Figure 2. The single stage protocol

The strength of this protocol lies in keeping the value of $\theta$ secret except for the communicating parties. The initial value of $\theta$ should be shared in a secure manner. In [8], the author has proposed using the three stage protocol to initially share the value of $\theta$. The initial $\theta$ will be used to transmit $k + n$ bits to Bob. From the transmitted bits, $n$ bits will be used to generate a new value of $\theta$. In this way, Alice and Bob derive the new $\theta$ from the transmitted bits and the initial $\theta$, and Eve cannot determine any information without knowing $\theta$.

## 3. CODING SCHEMES

The original implementations of BB84 and of the multi-photon tolerant quantum protocols have used the polarization of photons. Polarization coding is robust in free space optics because the medium does not depolarize the photons during transmission. However, polarization is not maintained in optical fibers because it may change due to external sources like stress, vibration, temperature, pressure or internal sources like impurities present in the core of the optical fibers. On the other hand, the pervasive presence of optical fibers (low losses, high speed and long transmission distance) in telecommunication networks has increased the interest in quantum communication using optical fibers. This necessitates the use of a polarization compensation mechanism to compute the drift and recover the original polarization. Passive and active polarization control systems have been designed to auto-compensate for the changes in the evolution of the state of polarization [9, 10, 11] as an optical beam travels over the fiber. However, no technique developed to date addresses long-haul QKD using polarization coded photons. Alternatives to polarization coding have been deployed in the past. The main alternative coding schemes are explained in the following subsections.

### 3.1. Time Coding

This subsection tackles the first alternative coding scheme: Time coding, which is better suited for fiber optic applications than polarization coding because it does not involve polarization mode dispersion and is more robust against medium losses. Time coding can be categorized in two protocols:

- Three Time Slots Protocol (3TS)

In 2002, Debuisschert and Boucher proposed using time coding with BB84 instead of polarization coding [12]. Alice sends a series of single photons to Bob with delay 0 (with respect to a reference) to encode bit 0 or delay $T/2$ to encode bit 1. Bob receives the pulses sent by Alice and sends at random half of them to an unbalanced Mach-Zehnder (MZ) interferometer to measure the pulse duration to detect possible shortening of pulses resent by Eve. The other half of the photons are sent to a photon counter to measure the detection time with respect to a reference to form the message sent by Alice. In this scheme, the binary information sequence is encoded on the time delay at the sender's side. The receiver establishes the sent sequence by measuring the time delay between the pulse and the reference.

- Coherent One Way (COW) QKD System

In 2005, the group of Applied Physics at the University of Geneva developed the COW QKD system [13]. Alice either generates coherent pulses with a mean photon number $\mu < 1$ or blocks the light completely (void state). Alice encodes the information using time slots separated by a time of $T$ and containing void or pulses with photons with the mean μ. Each bit is encoded in a sequence of two time slots. Decoy sequences are also used to keep the average number of photons at a certain limit to prevent Eve from changing the number of photons in the pulses without being detected. The $k^{th}$ logical bits 0 and 1 are encoded in a sequence of two states: a coherent state $|\sqrt{\mu}\rangle$ and a void state $|\sqrt{0}\rangle$ such that:

$$|0\rangle_k = |\sqrt{\mu}\rangle_{2k-1} |\sqrt{0}\rangle_{2k},$$
$$|1\rangle_k = |\sqrt{0}\rangle_{2k-1} |\sqrt{\mu}\rangle_{2k}.$$

The decoy sequence can be represented as $|\sqrt{\mu}\rangle_{2k-1} |\sqrt{\mu}\rangle_{2k}$. In this scheme, the logical bits are encoded on the sender's side using time slots, separated by $T$, containing void or $\mu$ pulses. The receiver, Bob, reveals for which bits he obtained detection in the data line and those detected in the monitoring line. Alice tells Bob which bits he has to remove from his raw key because they are due to the detections of decoy sequences. Then, Alice and Bob run error correction and privacy amplification ending up with a secret key.

## 3.2. Phase Coding

Phase coding, another alternative to polarization coding, is explained in this subsection. The phase of a photon is more stable than the polarization in optical fibers. The use of the phase of photons to encode information was first introduced by Bennett in 1992 [14]. The implementations of phase coding scheme can be divided into three categories [15]:

- Optical Fiber Version of MZ Interferometer

The optical fiber version of MZ interferometer consists of two couplers connected to each other with a phase modulator in each arm. Alice was represented by the light source, coupler and one phase modulator and Bob is another phase modulator, coupler and detectors.

- Double MZ Implementation

The optical fiber version of MZ interferometer works well in optical fiber medium but its problem lies in maintaining the stability of the path difference especially when Alice and Bob are kilometers away from each other. In 1992 as well, Bennett suggested a way to resolve this problem. He proposed using two unbalanced MZ interferometers connected in series by an optical fiber. Each MZ interferometer contained a phase modulator in the long arm in order to perform the encoding at Alice's side and the measurement at Bob's side.

- Plug and Play System

The "plug and play" system [16], proposed by Stucki et al. in 2002, refers to a system that does not require any optical adjustments. In the plug and play system, strong linearly polarized pulses travel from Bob to Alice. They are divided into two pulses P1 and P2. Alice applies a phase shift between the two pulses representing bit 0 or 1. This phase shift is then measured by Bob. Since compensation of polarization rotations in the fiber from Bob to Alice are performed on the way from Alice to Bob, this is an auto-compensating plug and play system.

## 3.3. Intensity Coding

In classical communication, amplitude modulation has been used to modulate a carrier signal to achieve communication over long distances. A similar approach can be applied in quantum cryptography [17]. From the quantum perspective, the intensity of a light wave is proportional to the square of the amplitude of the electric field of the light wave. At the sender's side, an intensity modulator will be used to encode the information by changing the intensity within the range $[I_i, I_f]$, i.e., perform intensity coding. At the receiver's side a second intensity modulator will be used to decode the information by reversing the transformation performed at the sender's side.

## 4. PROPOSED COMPACT CODING SCHEME

This section describes the general scheme and the steps for encoding and decoding information. In addition, this section defines the operations of the intensity modulator, the time delayer, and the phase modulator used in the proposed scheme.

## 4.1. Setup of the Proposed Scheme

The proposed compact coding scheme can be implemented using the setup shown in Fig. 3. A light source generating $n$ photons per pulse will be used. First, the light beam will pass through the intensity modulator where intensity coding is done. The resulting intensity modulated beam will pass through an unbalanced MZ interferometer with a time delayer in the long arm. Each pulse will be divided into two pulses with a time delay $\Delta + t$ added between them, where $\Delta$ is a fixed length of the time slots of the two pulses and $t$ is the encoded bit value in the form of a time delay within the slot. Now, the resulting beam is encoded both in intensity and time. Photon pulses now pass through a phase modulator which introduces a phase shift to them. The resulting intensity, time and phase encoded photon pulses are sent to Bob either over free space optics or optical fibers.

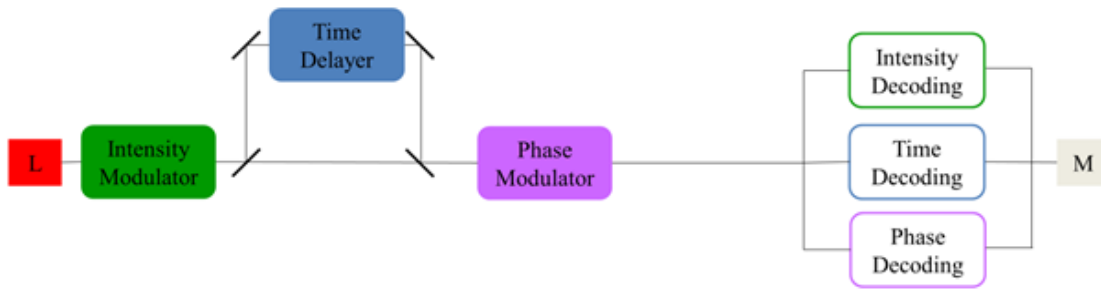

Figure 3. Schematic of compact coding

At Bob's side, the information carried by the pulses can in principle be decoded by dividing the $n$ photons in each pulse into three beams using a 1-3 beam splitter. The first path will perform intensity decoding, the second will undertake time decoding and the third will undergo phase decoding. The resulting measurements of all the three paths will be used to determine the encoded bits.

In the low-photon-number regime, to guarantee the security of the protocol [5, 6], it is more desirable to perform the decoding in a single channel, i.e., combing the three decoding paths into a single process (see Fig. 4). This can be accomplished for the case of binary coding in each of the modulations detailed in the following.

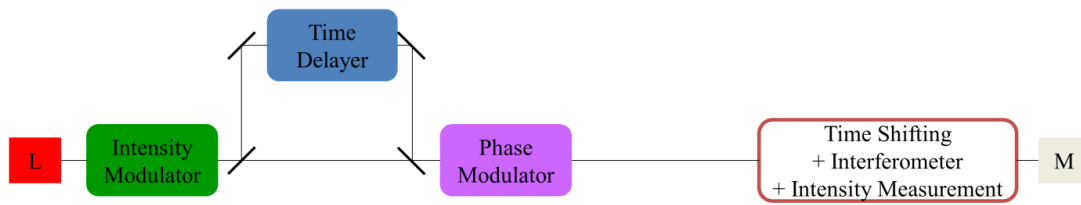

Figure 4. Schematic of simplified compact coding

## 4.2. Encoding and Decoding Example

As a basic example, assume there are only two values for each property of the pulse. Time can be $0$ or $T/2$, phase can be 0 or $\pi$ and intensity can be $I_1$ or $I_2$ (some constant values). The encoding stage proceeds according to Table 1 where each 3 bits of information correspond to a unique combination of intensity, time and phase values. Using this coding scheme, the field mode at the output of Alice's encoder can be written as

$$\frac{1}{\sqrt{2}}(a_0 + e^{i\phi}a_{\Delta+t}),$$

where $a_x$ denotes the pulse at time $x$ with $t = 0, T/2$ and $\phi = 0, \pi$. The intensity is proportional to $|a_x|^2$.

At the decoding stage, Bob first of all splits the pulses into two equal parts using a beam splitter. Since the two pulses $a_0$ and $a_{\Delta+t}$ are temporally separated, he can introduce a time delay $\Delta + t'$ to $a_0$ respectively to the output of the beam splitter, where $t' = 0, T/2$, giving $a_{\Delta+t'}$. This pulse is then combined with the pulse $a_{\Delta+t}$ using another beam splitter, resulting in

$$\frac{1}{2\sqrt{2}}(a_{\Delta+t'} \pm e^{i\phi}a_{\Delta+t}).$$

When $t' \neq t$, both the output ports of the beam splitter contain two pulses, i.e., $a_\Delta \pm e^{i\phi}a_{\Delta+T/2}$ or $a_{\Delta+T/2} \pm e^{i\phi}a_\Delta$. This scenario is not useful to decode the time and phase information. On the other hand, for $t' = t$, the output ports of the beam splitter are given by $(1 \pm e^{i\phi})a_{\Delta+t'}$, so that the time delay information is reflected in the two cases of $t'$ and the phase information is manifested in the bright and dark ports of the beam splitter. It should be noted that all the photons in both cases of $t' = t$ and $t' \neq t$ have to be measured, so that the total photon number gives the intensity $I_1$ or $I_2$. A schematic of the detection is shown in Fig. 5.

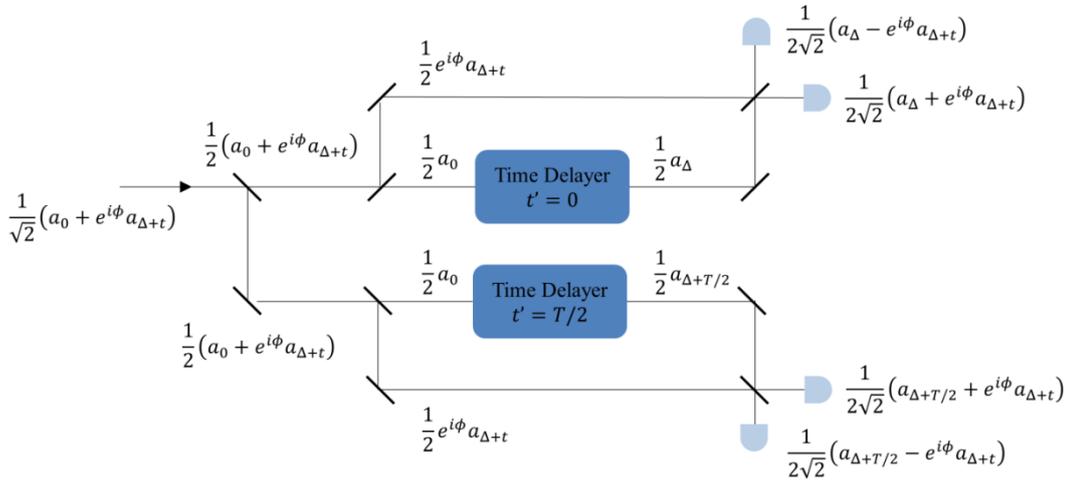

Figure 5. Schematic of combined time shifting, interferometer and intensity measurement for binary decoding in intensity, time and phase

The combination of the measured intensity, time and phase will be mapped to the corresponding bit combination in the Table 1. It should be noted that the configuration in Fig. 5 is not restricted to binary decoding for intensity and time, whereas the beam splitter needs to be generalized to a multi-port version for the phase discrimination if phase values other than $\phi = 0, \pi$ are used.

Table 1. Encoding and decoding example in compact coding.

| Bit | Intensity | Time ($t$) | Phase ($\phi$) |
|---|---|---|---|
| 000 | $I_1$ | 0 | 0 |
| 001 | $I_1$ | 0 | $\pi$ |
| 010 | $I_1$ | $T/2$ | 0 |
| 011 | $I_1$ | $T/2$ | $\pi$ |
| 100 | $I_2$ | 0 | 0 |
| 101 | $I_2$ | 0 | $\pi$ |
| 110 | $I_2$ | $T/2$ | 0 |
| 111 | $I_2$ | $T/2$ | $\pi$ |

## 4.3. Operations of Intensity Modulator, Time Delayer and Phase Modulator

In this subsection, the operations of the intensity modulator, the time delayer and the phase modulator used in the setup of the proposed scheme are described.

### 4.3.1. Operation of the Intensity Modulator

The intensity modulator will be used to change the intensity of the photons passing through it. The intensity of a light beam is proportional to the average number of photons in the beam. The larger the number of photons is, the higher the intensity of the light beam and vice versa.

A number state $|n\rangle$ represents a definite number of photons in the electromagnetic field. The number states are orthogonal and complete since

$$\langle n|m\rangle = \delta_{mn} \quad \text{and} \quad \sum_{n=0}^{\infty} |n\rangle\langle n| = 1, \quad \text{where } n = 0, 1, 2, \ldots.$$

$a$ and $a^\dagger$ represent the annihilation and creation operators of subtracting and adding a photon in a field composed of $n$ photons [18]. The application of the creation operator to a number state yields:

$$a^\dagger |n\rangle = \sqrt{n+1}|n+1\rangle.$$

The application of the annihilation operator to a number state yields:

$$a|n\rangle = \sqrt{n}|n-1\rangle.$$

Higher-order excited number states can be created from vacuum by repetitive application of the creation operator as follows:

$$|n\rangle = \frac{(a^\dagger)^n}{(n!)^{1/2}}|0\rangle.$$

These single photon operations remained a theory until 2004 when Grangier and coworkers implemented a simple way to subtract a photon from a traveling field using a beam splitter and a photodetector [19, 20]. In 2004 as well, Zavatta and coworkers performed an experiment to add a single photon to a coherent field [21]. In [22] the authors described the behavior of the beam splitter and how it can be used for photon addition and subtraction. If the traveling field is injected to the first input port while nothing is injected to the second input port, the action of photon subtraction is performed. If instead of the vacuum, a photon is injected in the second input port of the beam splitter and nothing is measured at the output, the action of photon addition is done. Another way for adding a photon uses a parametric down converter [22].

To increase the intensity of a light beam, a creation operator will be used to add $x$ photons. To decrease the intensity, an annihilation operator will be used to remove the $x$ added photons. The total number of photons, with the added or subtracted photons, should lie within the range $\{n_i, n_f\}$.

In the three stage protocol, a creation operator will be used to add the transformations corresponding to message encoding, Alice's transformation and Bob's transformation. Accordingly, an annihilation operator will be used to negate these transformations.

In the single stage protocol, a creation operator will be used to add the transformation corresponding to message encoding and a random transformation at Alice's side. At Bob's side, the removal of photons is replaced by a mathematical computation after the intensity of the received light beam is measured. A value corresponding to the added photons will be subtracted from the measured intensity [23]. Because the single stage protocol does not require removing photons at the receiving end, it is considered to be more secure than the three stage protocol from this perspective.

For the use of a weak laser beam represented by a coherent state

$$|\alpha\rangle = e^{-\frac{|\alpha|^2}{2}} \sum_{n=0}^{\infty} \frac{\alpha^n}{\sqrt{n}} |n\rangle,$$

the field amplitude $\alpha$ can be modulated by means of the displacement operator

$$D(\beta) = e^{\beta a^\dagger - \beta^* a},$$
$$D(\beta)|\alpha\rangle = |\alpha + \beta\rangle.$$

It is noted that the displacement operators commute with itself, i.e., $D(\alpha)D(\beta) = D(\beta)D(\alpha)$. This property makes the operation applicable to the three stage protocol.

**4.3.2. Operation of the Time Delayer**

The time delayer will be used to change the time between two pulses P1 (first pulse) and P2 (second pulse). To increase the time between the two pulses, a delay is added to P2 and P1 is kept intact as shown in Fig. 6. To decrease the time between the two pulses, P2 is kept intact and a delay is added to P1 as shown in Fig. 7.

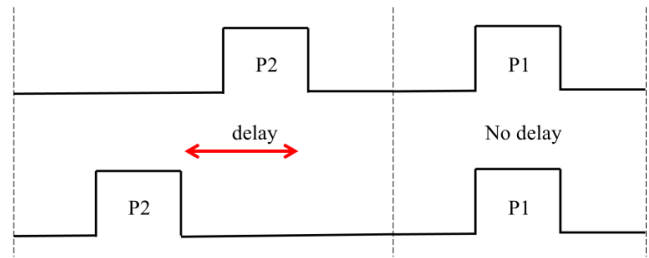

Figure 6. Time increase between two pulses P1 and P2

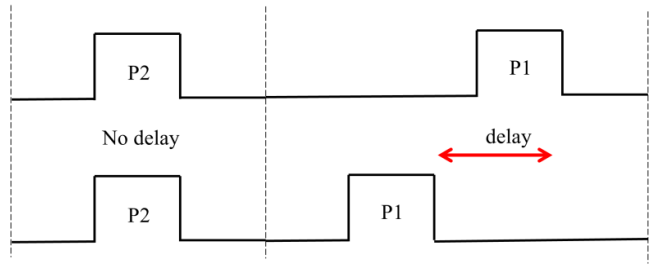

Figure 7. Time decrease between two pulses P1 and P2

The operation of the time delayer when the time between the two pulses is increased, i.e., a delay is introduced to P2, is illustrated in Fig. 8. Since the time delayer is synchronized with a clock, it can detect when the first pulse P1 arrives. The time delayer will be switched off. After a duration F, it will be switched on for a duration S and then switched off. The second pulse P2 can reach the time delayer anytime within this duration S and it will be subject to a time delay.

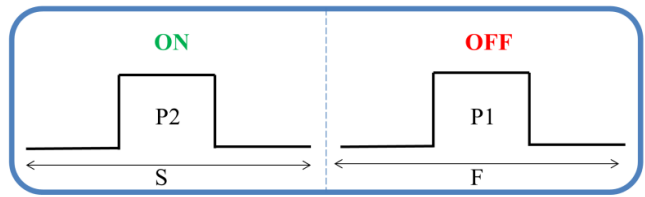

Figure 8. Operation of the time delayer for time increase

To negate the effect of the time delayer discussed in the previous paragraph, the time between the two pulses should be decreased by introducing a delay to P1 as illustrated in Fig. 9. Since the time delayer is synchronized with a clock, it can detect when the first pulse P1 arrives. The time delayer will be switched on for a duration F after which it will be switched off for a duration S. The first pulse P1 can reach the time delayer anytime within this F duration and it will be subject to a time delay. And the second pulse P2 will arrive to the time delayer within the S duration and will remain intact.

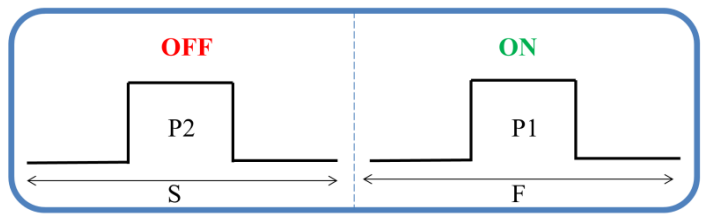

Figure 9. Operation of the time delayer for time decrease

### 4.3.3. Operation of the Phase Modulator

The phase modulator operates in a similar manner to the time delayer. The aim is to introduce a phase shift between the two pulses P1 and P2. Adding and negating the phase shift require the same technique: introducing some phase shift to the first pulse P1 and a different phase shift to the second pulse P2.

The phase modulator will be switching between two modes R and H. Mode R will produce a phase shift of $r$ and mode H will introduce a phase shift of $h$. The operation of the phase modulator is illustrated in Fig. 10.

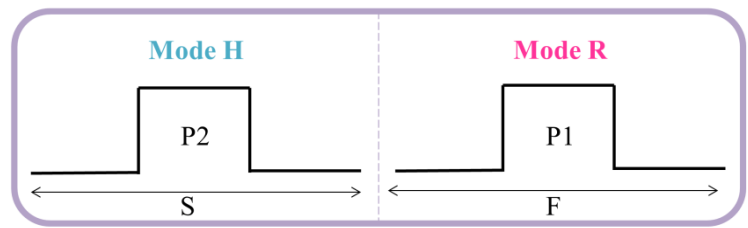

Figure 10. Operation of phase modulator

Synchronized with a clock, when P1 arrives within the F duration, the phase modulator will be operating in mode R. A phase shift of $r$ will be introduced to P1. Then, whenever P2 arrives within the S duration, the phase modulator will be operating in mode H. This introduces a phase shift of $h$ to P2. According to the choice of $r$ and $h$, the phase shift between P1 and P2 will either increase or decrease.

The mechanism of the phase modulation is similar to that of time coding, that is, by means of introducing a time delay with respect to a reference. The main difference is that, for the time coding, the time delay should keep the phase of the wavefront of the pulse fixed. The amount of delay is also chosen to be larger than the width of the pulse. On the other hand, the phase modulation is carried out by shifting the pulse such that the phase shift is within $2\pi$ with respect to the wavefront of the pulse. Finally, the durations S and F are usually taken to be the same, i.e., $S = F = \Delta$, for high speed operations in practice.

## 5. COMPACT CODING MULTI-PHOTON TOLERANT QUANTUM PROTOCOLS

This section proposes new protocols that result from the integration of compact coding scheme, proposed in the previous section, with the multi-photon tolerant quantum protocols. The structures of the proposed protocols, their steps and the key establishment procedures are described.

### 5.1. Compact Coding Three Stage Protocol

The structure of the proposed compact coding three stage protocol is sketched in Fig. 11. Alice has a message M, a bit string, which she wants to share with Bob. Alice is connected to Bob via a two-way quantum channel. The transmission medium can be free space optics or optical fibers.

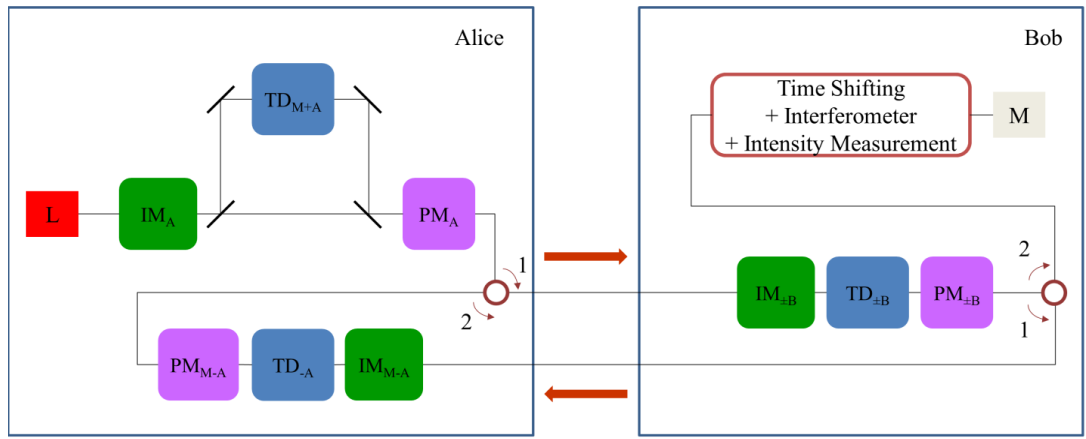

Figure 11. Schematic of the proposed compact coding three stage protocol

The proposed compact coding three stage protocol can be performed in four steps.

Step 1:

- A light source at Alice's side generates pulses each of $n$ photons.
- The intensity modulator applies a random intensity transformation ($I_A$) to the photons.
- The pulses of photons are split into two pulses using the MZ interferometer. P1 is the pulse that passes through the short arm of the MZ interferometer. The second pulse, P2, passes through the longer arm with a time delayer. The time delayer adds to P2 a delay corresponding to a random transformation applied by Alice ($D_A$) and a delay representing the encoded message ($D_M$).
- The two pulses P1 and P2 pass through a phase modulator that introduces a phase shift of $\alpha_A$ to P1 and $\beta_A$ to P2 corresponding to Alice's own transformation.
- Alice sends the resulting pulses to Bob.

Step 2:

- Bob receives the photon pulses and applies a random intensity transformation ($I_B$) to both pulses.
- The photon pulses pass through a time delayer. The time delayer keeps the delay of P1 intact and adds an additional delay ($D_B$) to P2.
- P1 and P2 pass through a phase modulator where a phase shift of $\alpha_B$ is added to P1 and $\beta_B$ is added to P2.
- Bob sends the pulses back to Alice.

Step 3:

- Alice negates the intensity transformation she applied in step 1 ($I_A$) and applies the transformation corresponding for encoding ($I_M$) to both pulses. The intensity transformation is negated using the method, utilized by the intensity modulator, explained in section 4.
- The pulses pass through a time delayer that adds a delay to P1 and keeps P2 intact. The delay added to P1 ($D_A$) corresponds to the delay introduced in step 1 to P2. Adding $D_A$ to P1 negates $D_A$ introduced to P2 as Alice's random transformation.

- P1 and P2 pass through the phase modulator that negates Alice's phase transformation $\alpha_A$ and $\beta_A$ from P1 and P2 respectively. A phase shift of $\alpha_M$ corresponding to encoding is introduced to P1.
- Alice sends the photons back to Bob.

Step 4:

- Bob negates the intensity transformation, time delay and phase modulation as what Alice does in Step 3.
- At this stage, the intensity, time, and phase information of the pulses should be identical to that Alice intended to send to Bob. Bob decodes this information by carrying the decoding method as described in section 4 to obtain the sent message M.

In the schematic diagram of key establishment procedure of the proposed compact coding three stage protocol (Fig. 12), $I_A$, $I_B$ and $I_M$ are the intensities for Alice's transformation, Bob's transformation and message encoding respectively. $D_A$, $D_B$ and $D_M$ are the time delays for Alice's transformation, Bob's transformation and message encoding respectively. ($\alpha_A$, $\beta_A$), ($\alpha_B$, $\beta_B$) and ($\alpha_M$, 0) are the phase shifts applied to pulses (P1, P2) for Alice's transformation, Bob's transformation and message encoding respectively.

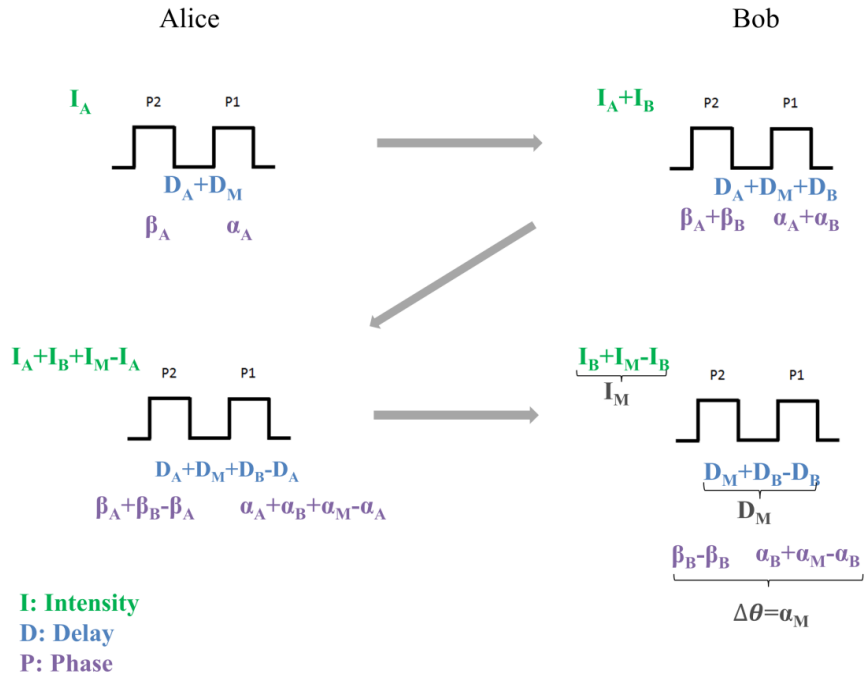

Figure 12. Key establishment procedure of the proposed compact coding three stage protocol

## 5.2. Compact Coding Single Stage Protocol

In the previous subsection, the compact coding three stage protocol was presented. This subsection proposes another protocol resulting from the integration of the compact coding scheme with another multi-photon tolerant quantum protocol, the single stage protocol.

Similar to the proposed compact coding three stage protocol, the proposed protocol in this subsection is also based on encoding the photon pulses using compact coding. The structure of the proposed compact coding single stage protocol is sketched in Fig. 13. Unlike the protocol in the previous subsection, communication between Alice and Bob in this structure involves sending the message only once instead of three times.

Alice and Bob have shared an initial value of $\theta$ in a secret manner ahead of time.

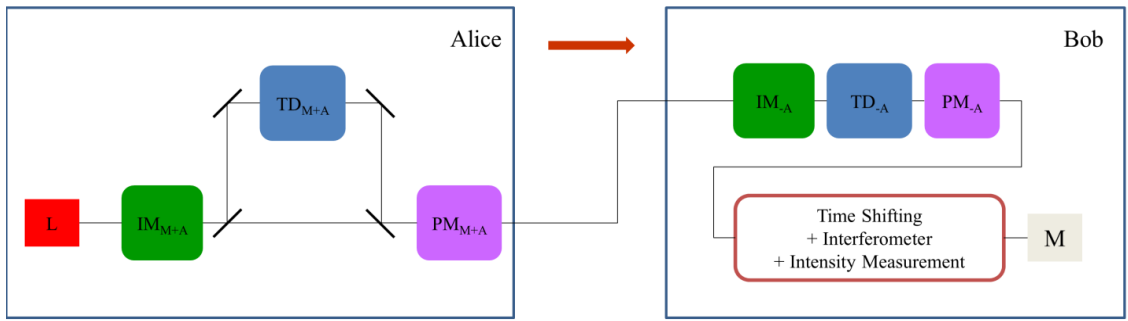

Figure 13. Schematic of the proposed compact coding single stage protocol

The proposed compact coding single stage protocol can be illustrated in three steps.

Step 1:

- A light source at Alice's side generates pulses each of $n$ photons.

- The intensity modulator applies to the photons the transformation corresponding for encoding the message ($I_M$) and another random transformation ($I_A$).
- The photon pulses pass through a MZ interferometer with a time delayer in the long arm. Each pulse is divided into two pulses P1 and P2. The time delayer keeps P1 intact and adds to P2 a delay corresponding to a random transformation ($D_A$) and a delay representing the encoded message ($D_M$).
- The two pulses pass through a phase modulator where a phase shift of $\alpha_A$ is added to P1 and $\beta_A$ is added to P2. A phase shift of $\alpha_M$ corresponding to encoding is introduced to P1 as well.
- Alice sends the resulting pulses to Bob.

Step 2:

- Bob negates the intensity transformation, time delay and phase modulation as what Alice does in Step 1.
- At this stage, the intensity, time, and phase information of the pulses should be identical to that Alice intended to send to Bob. Bob decodes this information by carrying the decoding method as described in the previous section to obtain the sent message M.

Step 3:

- With the help of a mathematical operation and part of the transmitted bit string, a new angle of transformation $\theta$ will be generated and used for the subsequent communication between Alice and Bob. This new value of $\theta$ will be plugged in the transformation formula $U_A$ and will be used in the next communication.

In the schematic diagram of key establishment procedure of the proposed compact coding single stage protocol (Fig. 14), $I_A$ and $I_M$ are the intensities for the transformation and message encoding respectively. $D_A$ and $D_M$ are the time delays for the transformation and message encoding respectively. $(\alpha_A, \beta_A)$ and $(\alpha_M, 0)$ are the phase shifts on pulses (P1, P2) for the transformation and message encoding respectively.

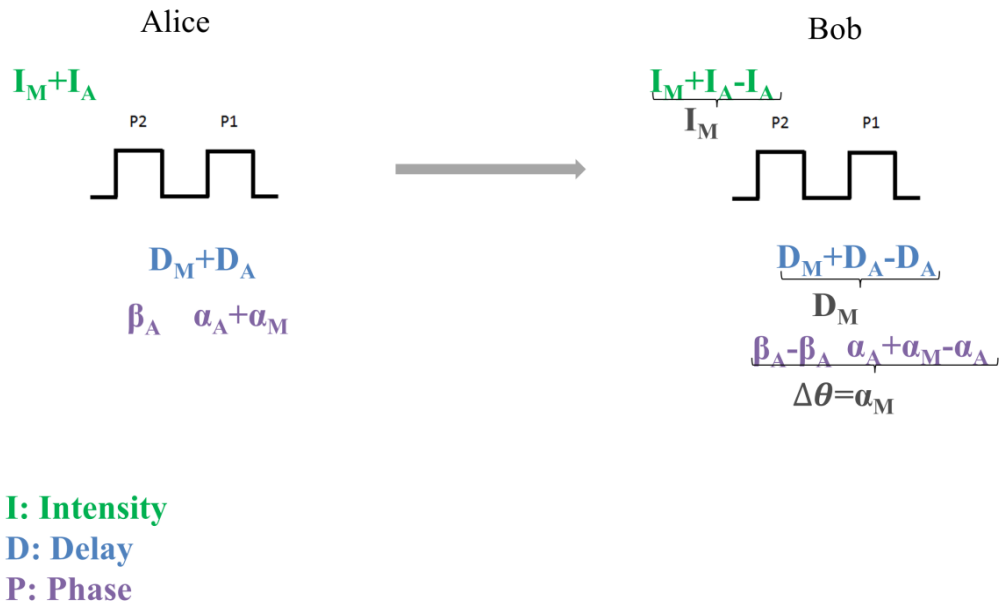

Figure 14. Key establishment procedure of the proposed compact coding single stage protocol

## 6. ANALYSIS AND FUTURE WORK

This section analyses the proposed protocols and sheds light on possible future work.

### 6.1. Analysis of the Proposed Protocols

The proposed compact coding scheme is an alternative to polarization coding which faces technical difficulties in practical implementation in optical fibers. It does not suffer from the problem resulting from uncontrolled polarization variation in optical fibers. Moreover, the application of the proposed coding scheme to multi-photon tolerant quantum protocols mitigates the problems of single photon generation, and distance and data rate limitations associated with single photon based protocols.

Another advantage of the proposed protocols is that Alice and Bob use different phases, time delays and intensities for each transmission due to the fact that the transformations change after each transmission. This prevents an intruder Eve from guessing the phases, time delays, intensities used and from even predicting the transformations used. Eve can apply the photon-number-splitting, man-in-the-middle, etc., as usual [6]. By doing that, she disturbs the information being sent and causes false detection at Bob's side. The use of phase shifts, time delays and intensity modulations add to the difficulty of Eve to infer the information being sent and hence increase her probability of being caught. The security of this system can be enhanced at the encoding level by first sending an initialization vector (IV) shared beforehand to check the presence of Eve before the initial transmission. In later transmissions, a part of the previously shared data will be sent to determine whether Eve is present or not. If Alice or Bob get any errors in the test key, IV or part of previously shared data, they know that Eve is present and the communication is stopped.

The security of the proposed protocols can be further improved by adding a fourth variable in the compact coding three stage protocol and in the three stage protocol preceding the compact coding single stage protocol. The aim is to thwart PNS attacks by making the system a four-variable three-equation system. The fourth variable is the IV shared beforehand for the first transmission. For later transmissions, the fourth variable will be the previously shared message M.

Because three properties of the photons are utilized for encoding: phase, time and intensity, each photon pulse can be used to encode at least three classical bits. Unlike other protocols that use only property, the proposed system can achieve three times higher data rate. It is worth noting that we are considering only two options for each property at the moment, i.e., the phase can be one of two values, and so do the time and the intensity. If we assign more than two options for each property, more bits can be encoded per pulse performing multi-level encoding and thus faster data rates can be achieved.

### 6.2. Future Work

The immediate steps that follow the proposed compact coding scheme are the hardware implementations of the general scheme, of the compact coding three stage protocol and of the compact coding single stage protocol. These implementations should be done over free space optics and optical fibers. The fiber implementations are expected to record higher data rates.
After the compact coding scheme in a multi-photon environment is implemented, the system can be modified by assigning more than two values for each property, i.e., transitioning from a binary system to an $m$-ary system. The relationship between the number of values for each property and the achieved data rate can be determined in order to maximize the rate at which data is transferred.

A further approach that can be taken is to test the robustness of each encoding property separately by sending the same message in the phase, time and intensity domains. The received messages in each property will be compared to determine which message incurred the least errors. Because the transmission channel introduces errors in the data, this approach enables the determination of the most robust property within the channel. Thus, more bits will be encoded using the most robust property and fewer bits will be encoded using the least robust property.

## 7. CONCLUSIONS

This paper has proposed a new encryption scheme *Compact Coding* that is applicable to either optical fibers or free space optics as the transmission medium. The proposed scheme utilizes for the first time three properties of the photons- phase, time and intensity- for secure information transfer simultaneously. It is also an alternative to polarization coding which has defied implementation in optical fibers. The structures of the proposed coding scheme in conjunction with multi-photon tolerant protocols have been presented and their operational details described. The application of the proposed coding scheme to multi-photon tolerant quantum protocols mitigates the problems of single photon generation, distance and data rate associated with single photon based protocols. Finally, the proposed compact coding multi-photon tolerant quantum protocols pave the way for a wide variety of future applications that require unconditionally secure data transfer. With the additional implementations, these protocols will resemble a promising solution for practical quantum communication.

## REFERENCES

[1] Singh S. (1999) The code book: the evolution of secrecy from Mary, Queen of Scots, to quantum cryptography, 1st ed., Doubleday, New York.
[2] The Heartbleed Bug. Available: http://heartbleed.com/
[3] Stallings W. (1999) Cryptography and network security: principles and practice, 5th ed., Prentice Hall, New Jersey.
[4] Kak S. (2006) "A three-stage quantum cryptography protocol," Foundations of Physics Letters, 19, 293-296.
[5] Chen Y.; Kak S.; Verma PK.; Macdonald G.; El Rifai M.; Punekar N. (2013) "Multi-photon tolerant secure quantum communication - From theory to practice," in IEEE International Conference on Communications (ICC), pp. 2111-2116.
[6] Chan KWC.; El Rifai M.; Verma PK.; Kak S.; Chen Y. (2015) "Multi-photon quantum key distribution based on double-lock encryption," in CLEO Conference, California, arXiv:1503.05793 [quant-ph].
[7] Thomas JH. (2007) "Variations on Kak's three stage quantum cryptography protocol," arXiv:0706.2888 [cs.CR].
[8] Darunkar B.; Verma PK. (2014) "The braided single-stage protocol for quantum secure communication," in Proc. SPIE 9123, Quantum Information and Computation XII, 912308.
[9] MacDonald G. (2003) "A polarization based intrusion monitoring system for optical networks," M.S. thesis, The University of Oklahoma, Oklahoma.
[10] Xavier G.; Walenta N.; Vilela de Faria G.; Temporão G.; Gisin N.; Zbinden H.; Von der Weid JP. (2009) "Experimental polarization encoded quantum key distribution over optical fibres with real-time continuous birefringence compensation," New Journal of Physics, 11, 045015.
[11] Ma L.; Xu H.; Tang X. (2006) "Polarization recovery and auto-compensation in quantum key distribution network," in Proc. SPIE 6305, Quantum Communications and Quantum Imaging IV, 630513.
[12] Debuisschert T.; Boucher W. (2004) "Time coding protocols for quantum key distribution," Phys. Rev. A, 70, 042306.
[13] Stucki D.; Brunner N.; Gisin N.; Scarani V.; Zbinden H. (2005) "Fast and simple one-way quantum key distribution," Appl. Phys. Lett. 87, 194108.
[14] Bennett C.; Bessette F.; Brassard G.; Salvail L.; Smolin J. (1992) "Experimental quantum cryptography," Journal of Cryptology, 5, 3-28.
[15] Gisin N.; Ribordy G.; Tittel W.; Zbinden H. (2002) "Quantum cryptography," Rev. Mod. Phys. 74, 145-195.
[16] Stucki D.; Gisin N.; Guinnard O.; Ribordy G.; Zbinden H. (2002) "Quantum key distribution over 67 km with a plug & play system," New Journal of Physics, 41.
[17] Hirota O.; Sohma M.; Fuse M.; Kato K. (2005) "Quantum stream cipher by the Yuen 2000 protocol: Design and experiment by an intensity-modulation scheme," Phys. Rev. A, 72.

[18] Parigi V.; Zavatta A.; Kim M.; Bellini M. (2007) "Probing quantum commutation rules by addition and subtraction of single photons to/from a light field," Science, 317, 1890-1893.
[19] Wenger J.; Tualle-Brouri R.; Grangier P. (2004) "Non-Gaussian statistics from individual pulses of squeezed light," Phys. Rev. Lett. 92, 153601.
[20] Namekata N.; Takahashi Y.; Fujii G.; Fukuda D.; Kurimura S.; Inoue S. (2010) "Non-Gaussian operation based on photon subtraction using a photon-number-resolving detector at a telecommunications wavelength," Nature Photon, 4, 655-660.
[21] Zavatta A.; Viciani S.; Bellini M. (2004) "Quantum-to-classical transition with single-photon-added coherent states of light," Science, 306, 660-662.
[22] Kim MS. (2009) "Quantum mechanics of adding and subtracting a photon," in Proc. SPIE 7236, Quantum Communications Realized II, 723605.
[23] Zavatta A.; Parigi V.; Kim MS.; Bellini M. (2008) "Subtracting photons from arbitrary light fields: experimental test of coherent state invariance by single-photon annihilation," New J. Phys. 10, 123006.

**AUTHORS**

**Rasha El Hajj** holds a Master's degree in Telecommunications Engineering from the University of Oklahoma. Her research interests are in the field of Quantum Cryptography and Data Encryption. In April 2014, she led the university's team to winning the first prize at a telecommunications-related business case competition held by Information and Telecommunications Education and Research Association in Louisville, Kentucky, USA.

**Pramode Verma** is Professor and Director of the Telecommunications Engineering Program in the School of Electrical and Computer Engineering of the University of Oklahoma-Tulsa. He also holds the Williams Chair in Telecommunications Networking. Prior to joining the University of Oklahoma in 1999 as the founder-director of a graduate program in Telecommunications Engineering, Dr. Verma held a variety of professional, managerial and leadership positions in the telecommunications industry at AT&T Bell Laboratories and Lucent Technologies.

**Kam Wai Clifford Chan** is an assistant professor of the Telecommunications Engineering Program in the School of Electrical and Computer Engineering, University of Oklahoma-Tulsa. He received the Ph.D. degree in physics from the University of Rochester in 2005. Prior to joining the University of Oklahoma, he was a research scientist at Rochester Optical. His research interests include quantum optics, quantum imaging, and quantum communications.